\documentclass[12pt]{article}
\usepackage{dsfont}
\usepackage{latexsym,amsfonts,amsmath,amssymb,mathrsfs}
\usepackage{verbatim,cite}
\usepackage{graphicx}
\usepackage{floatflt}
\usepackage{subfigure}
\usepackage{color}
\usepackage{setspace}
\usepackage{wrapfig}
\usepackage{empheq}
\usepackage[weather]{ifsym}
\usepackage{stmaryrd}
\usepackage{stackengine}
\usepackage{picins}
\usepackage{lipsum}
\usepackage[multiple]{footmisc}
\usepackage[export]{adjustbox}
\usepackage{eurosym}
\usepackage{mathptmx}
\usepackage{enumitem}
\usepackage{fancyhdr}
\usepackage{mathtools}
\usepackage{physics} 
\usepackage{quantikz}

\def\c1{{\textcircled{a}}}

\def\@begintheorem#1#2{\tmpitemindent\itemindent\topsep 0pt\rm\trivlist
    \item[\hskip \labelsep{\indent\it #1\ #2:}]\itemindent\tmpitemindent}
\def\@opargbegintheorem#1#2#3{\tmpitemindent\itemindent\topsep 0pt\rm \trivlist
    \item[\hskip\labelsep{\indent\it #1\ #2\
    \rm(#3):}]\itemindent\tmpitemindent}
\def\@endtheorem{\endtrivlist\unskip}

\def\0{{\mathbb 0}}

\makeatletter \renewcommand\d[1]{\ensuremath{%
  \;\mathrm{d}#1\@ifnextchar\d{\!}{}}}
\makeatother

\def\XXint#1#2#3{{\setbox0=\hbox{$#1{#2#3}{\int}$ }
\vcenter{\hbox{$#2#3$ }}\kern-.6\wd0}}

\begin{document}

 \setcounter{page}{1}

\begin{center}
\textbf{\Large Quantum Compressive Sensing Meets Quantum Noise: \\ A Practical Exploration}
\end{center}
\begin{center}
\textit{Naveed Naimipour$^{1,2}$, Collin Frink$^{ 2,3}$, Harry Shaw$^{ 2}$, Haleh Safavi$^{2}$, and Mojtaba Soltanalian$^{1}$}
\end{center}
\begin{center}
$^{1}$ Department of Electrical and Computer Engineering, University of Illinois Chicago, Chicago, IL 60607, USA \\
{$^{2}$ NASA Goddard Space Flight Center, Greenbelt, MD 20771, USA \\
$^{3}$ Peraton, Reston, VA 20190, USA}
\end{center}

\vspace{-0.36in}

\section{Abstract} \vspace{-0.18in}

Compressive sensing is a signal processing technique that enables the reconstruction of sparse signals from a limited number of measurements, leveraging the signal's inherent sparsity to facilitate efficient recovery. Recent works on the Quantum Compressive Sensing (QCS) architecture, a quantum data-driven approach to compressive sensing where the state of the tensor network is represented by a quantum state over a set of entangled qubits, have shown promise in advancing quantum data-driven methods for compressive sensing. However, the QCS framework has remained largely untested on quantum computing resources or in the presence of quantum noise. In this work, we present a practical implementation of QCS on Amazon Braket, utilizing the Quantum Imaginary Time Evolution (QITE) projection technique to assess the framework's capabilities under quantum noise. We outline the necessary modifications to the QCS framework for deployment on Amazon Braket, followed by results under four types of quantum noise. Finally, we discuss potential long-term directions aimed at unlocking the full potential of quantum compressive sensing for applications such as signal recovery and image processing.

\vspace{-0.22in}

\section{Introduction} \vspace{-0.18in}

As practical quantum systems continue to grow in complexity and scale, quantum noise has become a critical obstacle, significantly affecting quantum operations and measurements. Quantum noise encompasses various types of disturbances that impact quantum systems, each influencing the accuracy and stability of quantum states. One of the most well-studied types of noise is axis-flip noise, which occurs when the measurement axis flips, and is commonly seen in quantum systems where the state is represented on the Bloch sphere. Axis-flip noise causes the state to randomly rotate, leading to errors in the system's evolution \cite{Nielsen2012, axisflip, axisflip2, axisflip3}. Similarly, depolarizing noise occurs when a quantum state randomly transforms into a completely mixed state, reducing the system's coherence. This noise diminishes the information content of quantum states, making it a major concern in quantum error correction \cite{depolarizing, depolar, depolar2, depolar3}. It should be noted that quantum decoherence, resulting from shot noise, amplitude damping, and thermal noise, can also significantly impact signal reconstruction, making it even more difficult to analyze \cite{noiseSystem, NISQnoise, noiseSpec}.

Shot noise arises due to the probabilistic nature of individual particle movement, such as photons or electrons, when they are detected in a system. This statistical fluctuation occurs because of the random arrival times and quantities of these particles over a finite period, leading to uncertainty in measurements, especially in systems that rely on counting discrete events. Shot noise becomes particularly significant in low-intensity measurements, such as those in quantum optics, where precision is critical \cite{shotNoise, shotnoise2, shotnoise3}. Additionally, amplitude damping, resulting from the decay of energy or population of quantum states, presents a continuous challenge within quantum systems \cite{ampdamp, Ampdamp2, circamplitude}. Finally, thermal noise arises when the environment is at higher temperatures, becoming an increasingly larger issue as quantum systems continue to scale in size \cite{thermalnoise, thermalnoise2, thermalnoise3, datatransthermal}. Together, these noise types set fundamental limits on the performance of quantum technologies, particularly in quantum computing and quantum communication, where maintaining stable, coherent states is crucial for successful operation and error-free information transfer \cite{NISQnoise, datatransthermal}. In order to combat these limitations, a numbers of approaches have been developed to utilize powerful machine learning techniques to model the effects of quantum noise and decoherence with the hope of future development on noisy quantum systems \cite{Abbas_2021, Biamonte_2017, Bittel_2021, Huang_2021}.

 Within the context of quantum computing, machine learning techniques, or architectures inspired by them, have had difficulty dealing with quantum noise in practical applications \cite{NISQnoise, Huang_2021}. For example, although quantum autoencoders have been shown to be effective in developing noise reduction schemes, utilizing machine learning autoencoders for further performance enhancement has been seen as a long term goal due to complexity and hardware limitations \cite{autoencQ, QautoEnc}. An alternative example lies within techniques such as reinforcement learning and graph neural networks, which have been utilized to enhance quantum error correction (QEC) protocols \cite{qec_proc, Liao_2024, AndreassonRL_QEC}. This has been shown to guide the correction process, improving the overall accuracy of quantum systems in noisy environments. Furthermore, robust training methodologies rooted in adversarial training or data augmentation have shown mixed results when attempting to improve the resilience of machine learning models to quantum noise \cite{Huang_adver, liaoQuant}. Despite these significant advancements, there remains a clear gap in existing research in relation to how these systems or techniques would perform with practical data or architectures. The area of compressive sensing, which tackles the problem of signal reconstruction, offers a unique lens into this problem.

Compressive sensing is a signal processing technique that offers an intriguing method for reconstructing signals. Mathematically, the measurement procedure is described by a sensing matrix $A$ relating the measurement vector $x \in \mathbb{R}^m $ and the signal vector $ y \in \mathbb{R}^n $ via  $x = A y $, where $ m \ll n $ and the signal $y$ must be sparse to enable reconstruction. This is achieved by minimizing the  $\ell_0$ -norm of all possible signals while ensuring consistency with the measurements. The traditional approach of sampling a signal at regular intervals, as dictated by the Nyquist-Shannon sampling theorem, becomes inefficient when dealing with sparse or structured signals \cite{CS1, CANDES_RIP, CS4, CS2, CS3}. Compressive sensing addresses this inefficiency by utilizing the sparsity of the signal \cite{CS3, candes_sparsity}. Specifically, rather than measuring all components, it takes fewer strategically chosen measurements that can later be used to reconstruct the signal \cite{CS2, CS3, candes_sparsity}. This reconstruction can be achieved through optimization methods that exploit sparsity, using fewer measurements than conventional methods would require and has resulted in compressive sensing being found in everything from underwater imaging to information security \cite{candes_06, candes_06_2, Candes_2007, underwater_image, security1, security2}. 

The practical implementation of compressive sensing was shown via a single-pixel camera and a micro-array of mirrors \cite{Duarte_2008}. The mirrors were randomly configured to direct light into or away from the camera and each measurement corresponded to the total intensity of light for a specific mirror configuration, a process equivalent to integrating over a random selection of pixels in a conventional CCD image. This was implemented in LIDAR geoimaging, which can be utilized to improve the accuracy and efficiency of data analysis and modeling in various Earth science domains such as climate models, refining geological surveys, and advancing remote sensing technologies \cite{Howland_2011}.

The idea of quantum compressive sensing emerged as a natural extension of classical compressive sensing, leveraging quantum systems' unique properties to provide even more efficient data sampling and reconstruction with the help of machine learning. This is particularly illustrated with the adaptation of a ``Boltzmann machine" into a ``Born machine," which represents the probability distributions using complex-valued amplitudes whose square moduli correspond to the same distribution as that of the Boltzmann machine\cite{Wang_boltzmannVSborn}. In this context, a Born machine can be understood as a quantum state realized with qubits, where entanglement captures correlations in the data \cite{Wang_boltzmannVSborn, BornMachine1}. 

Tensor network-based methods have emerged as practical alternatives for large scale quantum computing systems as they continue to develop. The Matrix Product State (MPS) tensor network, which approximates quantum states efficiently, has been applied to machine learning tasks such as supervised classification, generative modeling and Tensor Network Compressed Sensing (TNCS) \cite{Stoudenmire_2016, Han_2018, Ran_2020}. Furthermore, architectures such as Quantum Compressive Sensing (QCS), which was inspired by TNCS, illustrated how data-driven approaches such as tensor networks could be trained to learn the structure of signals and then updated to align with the measured data before being sampled to estimate the original signal \cite{QCS}. QCS proposed a promising alternative ``quantum" protocol, where the tensor network's state is represented as a quantum state over a set of entangled qubits, leveraging quantum theory to enhance the signal reconstruction process in noiseless scenarios with LIDAR data \cite{QCS}. 

In this paper, we continue the development of the Quantum Compressive Sensing architecture by exploring its ability to handle quantum noise when dealing with practical data. Specifically, we utilize the protocol to implement one of the more promising projection algorithms that was proposed in it's full form, quantum imaginary time evolution (QITE) algorithm, on Amazon's quantum cloud computing services for a practical LIDAR dataset in the presence of quantum noise \cite{Motta_2020FEB, braket}. We then discuss the results and explore long term possible directions to continue expanding on the idea of a practical quantum compressive sensing architecture.

\section{Methodology} 
\vspace{-0.18in}
In this section, we will delve into the specifics of our (QCS) framework \cite{QCS}, the quantum noise considered for implementation, and the extensions necessary for a meaningful results on quantum computing resources.

\subsection{Quantum Compressive Sensing: A Brief Summary}
The original Quantum Compressive Sensing protocol can be summarized in four stages --- Training, Measurement, Projection, Sampling. Although each stage serves a distinct purpose theoretically, the framework was not completely optimized for a practical implementation. For the purposes of this work, we will focus on the relevant details that are necessary for the presented implementation and refer to \cite{QCS} for more detailed explanations of each stage. 

The training stage prepares an $n$-qubit Born machine $\ket{\Psi}$ to represent the probability distribution of signals $y$, mapping each signal to a complex amplitude $\braket{y}{\Psi}$ whose square modulus reflects the probability of observing the signal $y$. Furthermore, the measurement stage performs $m$ distinct measurements on an unknown $n$-dimensional signal $y$ using a $m \times n$ sensing matrix, with outcomes given by $x = Ay$. Within the projection stage, the Born machine $\ket{\Psi}$ is projected onto a state $\ket{\Psi_x}$ that aligns with the measurement outcome $x$, involving complex operations with a non-unitary operator. It should be noted that for this stage, both $A$ and $x$ are considered as inputs in the projection stage. Finally, qubits are measured to produce a binary string representing an estimated signal $y'$ in the sampling stage. The reconstructed signal $y'$ is an approximation if the original signal $y$ is non-binary, but this is typically sufficient to convey the relevant data. Furthermore, the sampling process can be repeated multiple times to generate a sequence of binary signals, whose statistics could be mapped back to the non-binary image. However, since both the projection and sampling steps alter the machine's state, the Born machine $\ket{\Psi}$ must be retrained after each sample.

\subsubsection{Data Mapping}

One of the challenges with any quantum protocol for practical applications is devising an approach to map pixels to qubits, typically involving encoding a signal into a quantum state. In the case of QCS, a pixel-qubit mapping is proposed where a signal $ y $ is represented as an $ n $-dimensional vector $ y \in Y \subset \mathbb{R}^n $, where $ Y $ denotes the set of all possible signals that are expected be measured. The goal is to prepare a Born machine $ |\Psi\rangle $ based on the set $ Y $, which will be implemented as a quantum state of $ n $ qubits.

To achieve this, a method is needed to map a signal $ y \in Y $ to a quantum state $ |y\rangle $, which is an element of the $ d^n $-dimensional Hilbert space associated with $ n $ qudits. The focus of this work will be similar to \cite{QCS}, meaning we take the case $d = 2$ (the standard ``qubit") and $ n' = n $. In other words, we will focus on the case where one qubit will be used for each ``pixel" in our image. This approach enables $ |y\rangle $ to be represented as an entanglement-free product state, where the state of each qubit is independently determined by the corresponding pixel value $ y_i $.

For binary signals, the natural approach is to map them to the computational basis states. For real-valued pixels, they will be mapped to a linear combination of these two basis states. \cite{Stoudenmire_2016} provided an example of this mapping technique that was further generalized by QCS \cite{QCS} to designate an arbitrary pixel value $v$ other than $0.5$ as the ``midpoint'' between $0$ and $1$:

\begin{align}
\label{eq:mapping:scaled2}
    \ket{y_i} &\rightarrow \cos\qty[\frac{\pi}{2}f_v(y_i)] \ket{0} + \sin\qty[\frac{\pi}{2}f_v(y_i)] \ket{1},
\end{align}
where the function $f_v$ is given by
\begin{align}
\label{eq:fpx2}
    f_v(x) &= \frac{1}{2} \qty[1 + \frac{2}{\pi} \arctan\qty[\tan(\pi\qty(x-\frac{1}{2})) - \tan(\pi\qty(v-\frac{1}{2}))] ].
\end{align}
The function $f_v: \qty[0,1]\rightarrow\qty[0,1]$ reduces to $f_v(x)=x$ for $v=0.5$, and becomes a smooth non-linear rescaling of $y_i$ before applying Equation~\ref{eq:mapping:scaled2} for $v\ne0.5$.
As a result, the state specified by Equation~\ref{eq:mapping:scaled2} for an arbitrary $n$-dimensional signal $y$ is a product state efficiently prepared by $n$ parallel single-qubit gates.

\subsubsection{Training}
For training, QCS \cite{QCS} proposes a novel approach that avoids gradient descent or numerical optimization methods to prevent the commonly encountered convergence difficulties. The goal is to prepare $n$ qubits into a Born machine $\ket{\Psi}$ describing the set $Y$. With this in mind, QCS illustrated that the Born machine state $\ket{\Psi}$ can be modeled as a ``quantum average" of elements in a set $Y$ as in \cite{Ran_2020}, which can be described as the superposition of each element in $Y$:

\begin{align}
\label{eq:quantum:average2}
    \ket{\Psi} = \frac{1}{\sqrt{|Y|}} \sum_{z=0}^{|Y|-1} \ket{y_z}.
\end{align}

If all elements in $Y$ are orthogonal and $y$ is uniformly distributed, the squared amplitude $\qty|\braket{y}{\Psi}|^2 = \frac{1}{|Y|}$, matching the desired probability of retrieving $y$ from $Y$. For non-orthogonal elements, useful results can still be achieved by leveraging overlaps between similar images, constructing $\ket{\Psi}$ from a training set $D$.

While no straightforward quantum circuit prepares a sum of non-orthogonal states deterministically, QCS devises a quantum circuit that prepares the following state:

\begin{align}
\label{eq:quantum:average:circuit}
    \ket{\Psi} = \frac{1}{\sqrt{|D|}} \sum_{z=0}^{|D|-1} \ket{z} \sum_{z'=0}^{|D|-1} (-1)^{z \cdot z'} \ket{y_{z'}},
\end{align}
where $\ket{z}$ represents the control register's computational basis states with $\log_2 \qty|D|$ qubits, and $\ket{y_z} \equiv U_z \ket{\vb{0}}$ is the state-preparation circuits using the Pauli $Y$ spin operator. We will have prepared the desired linear combination as described by Equation~\ref{eq:quantum:average2} if we measure all the control qubits in the basis state $0$. The reasoning behind this approach is illustrated by the probability of measuring all control qubits in the state $0$, utilizing the ``reduced density matrix'' of the control register (obtained by tracing out all degrees of freedom in the state register). In short, the aforementioned probability when utilizing the qubit mapping described in \ref{eq:mapping:scaled2} will be $\ge1/\qty|D|$, meaning the circuit describing Equation~\ref{eq:quantum:average2} has the advantage of being executed only $O(\qty|D|)$ times in order to successfully prepare the quantum-average state.

\subsubsection{Projection}
\label{sec:projectionQITE}
Non-unitary transformations are challenging to implement on a quantum computer, making the projection step, along with the large number of qubits required, a significant obstacle for implementing works like TNCS. The approach presented by QCS to achieve this projection involves measuring $m$ of the $n$ qubits to collapse them to basis states, with entanglement ensuring that unmeasured qubits also collapse accordingly. However, to prepare the desired state $\ket{\Psi_x}$, the training and measurement steps must be repeated until the measured qubits' values align with the classical measurement outcome $x$. With that in mind, three projection techniques were introduced --- Decomposition, Rodeo Algorithm, and Quantum Imaginary Time Evolution. 

The decomposition technique, as its name suggests, uses singular-value decomposition of the sensing matrix $A$. This method is a meaningful generalization in theory, but its primary limitation arises from the fact that it is mostly suitable for a restricted class of sensing matrices. In particular, the rotation matrix in the decomposition must be a generalized permutation matrix, making meaningful practical results for arbitrary sensing matrices unlikely. The Rodeo technique, taking inspiration from the probabilistic method based on quantum phase estimation originally developed for eigenvalue identification of a Hamiltonian operator \cite{Choi_2021}, involves applying a time-evolution operator to a trained Born machine, with a random time $\tau$. The phase induced is then transferred to a control register, and after a Quantum Fourier Transform, can be measured by inspecting all qubits. Although this process allows for extracting phase information efficiently, success of the algorithm depends on the phase measurement resolution, which is determined by the number of qubits in the control register. This creates a limitation, as higher resolution demands more qubits and increased qubit connectivity to manage controlled operations, making the protocol more resource-intensive. Finally, the Quantum Imaginary Time Evolution (QITE) technique introduces a Gaussian operator whose action on a computational basis state can then be defined and applied on the Born machine. As a result, this exponentially suppresses any basis state whose expectation value differs from the measurement outcome. Although the Gaussian operator is not unitary or idempotent, QITE can be used to implement a unitary operator that effectively achieves the same result for the specific input state. Since the work done in \cite{QCS} relied on classical simulation and the operator proposed was diagonal, the implementation of this projection technique side stepped the exponential scaling issues along with any complications arising from the many small time-steps of the Trotter decomposition in \cite{Motta_2020FEB}. However, QITE remains the most promising technique due to its design flexibility and proven statistical prowess. As a result, a full implementation of QITE, instead of the simplified version in QCS, would best display the capabilities of the QCS framework in its current form on quantum computing resources. 

\begin{figure}[h]
    \centering
    \includegraphics[width=0.9\columnwidth]{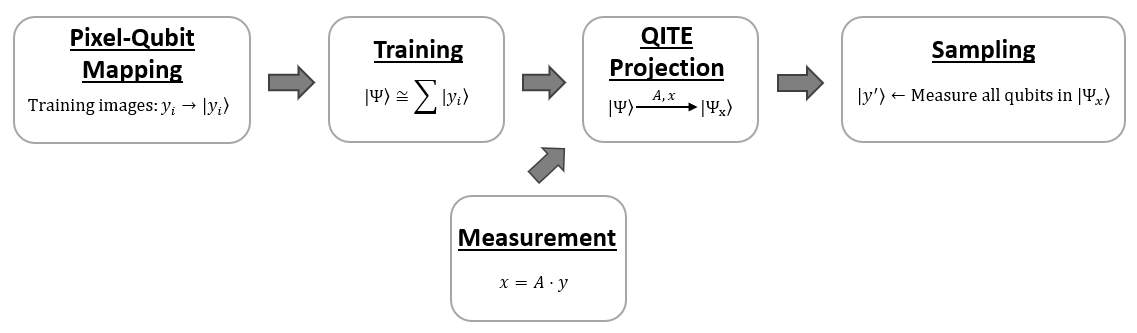}
    \caption{General workflow diagram of the proposed protocol where it can be seen the Pixel-Qubit mapping occurs before undertaking training. The QITE projection can then utilize the training and measurements which ultimately leads to the sampling.}
    \label{fig:workflow2}
\end{figure}

A brief review of the full QITE proposed in \cite{Motta_2020FEB}, and implemented in this work, begins with quantum state tomography over $\ket{\Psi}$ to construct an effective Hamiltonian 

\begin{align}
\label{eq:QITEhamil}
\hat H = \sum_{m=0}^{M} \hat h[m], 
\end{align}
where $\hat h[m]$ acts on a local set of qubits. The next step involves the implementation of the time-evolution 

\begin{align}
\label{eq:timeevol}
e^{- \beta \hat H }= \prod_{k} (e^{-\Delta\tau \hat H[k]})^s +  O(\Delta \tau), 
\end{align}
where $s=\frac{ \beta}{ \Delta \tau}$ is the number of imaginary time steps, over many small time-steps via Trotter decomposition to a state $\ket{\Psi}$. This full implementation within the QCS allows for a more meaningful results when tested with quantum noise. In the next section, we will describe the types of quantum noise that we will be focusing on. \figurename~\ref{fig:workflow2} shows a visual overview of the protocol as a whole for this work.

\vspace{-0.22in}
\subsection{Quantum Noise} 
\label{qnoise}
\vspace{-0.15in}

One of the primary challenges in realizing quantum computing hardware is maintaining quantum states for a long enough period to perform the desired computations. Given that near-term quantum computers are especially prone to noise effects, it is essential to study how these effects impact the performance of frameworks such QCS if they are to be developed into practical architectures long term. Although real-world noise models are beyond the scope of this work, three foundational computational representations of noise are described below and will be integrated into the QCS LIDAR results.

\subsubsection{Axis-Flip Noise}
\label{flipnoise}

Probabilistic axis-flip noise models are one of the initial steps toward the long-term goal of implementing real-world noise models within quantum algorithms. The rationale of this approach is that axis-flips are among the simplest modifications of a quantum state to understand. Since they can be implemented in both statevector and density-matrix simulations, it is computationally feasible to incorporate axis-flip noise models in relatively large simulations compared to other types of noise. While there may not be a direct physical process motivating the study of types of axis-flip noise, it remains an important building block in this process.

A bit-flip in the context of quantum algorithms, assuming Z-basis, is a switch from the $\ket{0}$ to the $\ket{1}$ state. For one qubit, it is implemented by a Pauli X-gate:

\begin{equation}
    \sigma_X = \begin{bmatrix}
        0 & 1\\
        1 & 0 
    \end{bmatrix} .
\end{equation}

Generally, if bit-flips are manually inserted into the code and the bit-flips are applied with probability $p$ after each gate layer. If working with a density matrix $\rho$, the state $\rho'$ resulting from a bit-flip on qubit $i$ is:

\begin{equation}
    \rho ' = (1-p)\rho + p X \rho X^\dag . 
\end{equation}

 It is worth noting that $\sigma_X$ can be easily substituted with $\sigma_Z$ and $\sigma_Y$ to generate phase-flip and bit-phase-flip noise, respectively. In particular, phase-flip noise (also termed dephasing noise), has been shown to map to a real-world quantum phenomena of phase loss in various types of qubits \cite{Nielsen2012}.

\subsubsection{Depolarizing Noise}
\label{depolarizing}

Depolarizing noise is another crucial element to study when understanding real-world noise effects. Unlike axis-flip noise, which modifies the quantum state with a certain probability, depolarizing noise completely destroys our knowledge of the quantum state. In the density matrix notation for a single qubit, this is represented as:

\begin{equation}
    \rho ' = (1-p)\rho + \frac{p}{2}I , 
\end{equation}
where $I$ is the I identity matrix. Fortunately, this noise is easily approximated on statevector simulations using a similar approach to that in section 3.4.2 by noting \cite{Nielsen2012}:

\begin{equation}
    \frac{I}{2} = \frac{\rho + \sigma_X \rho \sigma_X + \sigma_Y \rho \sigma_Y + \sigma_Z \rho \sigma_Z}{4}.
\end{equation}

Instead of only using $\sigma_X$ for bit-flips as in Subsection \ref{flipnoise}, by randomly selecting between the Pauli $\sigma_X$, $\sigma_Y$, and $\sigma_Z$ operators we are able to statistically model depolarizing noise assuming a sufficiently high number of simulation trials.

\subsubsection{Amplitude Damping} \label{amplitudedamping}
  One shortcoming of both axis-flip and depolarizing noise is the inherent symmetry between the states $\ket{0}$ and $\ket{1}$. However, in many quantum hardware schemes, this symmetry does not hold many real world noise effects \cite{Nielsen2012}. For example, with amplitude damping noise, the state $\ket{1}$ decays into $\ket{0}$ \cite{Nielsen2012}. The density matrix representation of this noise effect is:

\begin{equation}
    \rho ' = K_0 \rho K_0^\dag + K_0 \rho K_0^\dag ,
\end{equation}
where 
\begin{equation}
    K_0 = \begin{bmatrix}
        1 & 0\\
        0 & \sqrt{1-d}
    \end{bmatrix},
\end{equation}

\begin{equation}
    K_1 = \begin{bmatrix}
        0 & \sqrt{d}\\
        0 & 0
    \end{bmatrix} .
\end{equation}

It should be noted that amplitude damping is difficult to efficiently integrate into statevector simulations. As a result, we limit the extent of our results to the density matrix simulations of noisy QITE.

\subsubsection{Shot Noise}
\label{shotnoise}

The final noise studied in this work is shot noise, stemming from the probabilistic nature of individual particle movement in the quantum system. One of the most fundamental ideas in quantum mechanics is that the wavefunction of a quantum state ``collapses" into one of its eigenstates upon observation of the state. However, many quantum algorithms, including QITE, necessitate the measurement of the partial or full probability distribution of the set of states in the wavefunction. In practice, this means that when choosing a finite number of shots, there must be a balance between the cost of each shot with the overall reduction in expected error to the true probability. There are numerous ways to model shot noise when testing on non-quantum systems, one of the most common being simulating random arrival times of particles via the Poisson distribution  \cite{shotNoise, shotnoise2, shotnoise3}. As our implementation will be on quantum cloud computing resources, this will be unnecessary as shot noise will naturally exist. Within the QCS pipeline, shot noise is significant in two places: the final estimate of  reconstructed portion of the signal and the intermediary steps of QITE projection, as seen in \figurename~\ref{fig:shotnoise}, and will be discussed in the results section. 

\subsection{Other Types of Noise}\label{othernoise}
We emphasize the noise statistics of large quantum computers are too complex and nuanced to be accurately modeled by axis-flip, decoherence, or amplitude damping models alone. It is necessary to tailor the parameters of real-world noise models to the specific hardware on which the algorithm will be implemented \cite{Georgopoulos2021}. Due to these complexities, this work focuses on the noise models discussed in Subsections~\ref{flipnoise}-\ref{shotnoise}. In the future, however, the specific application of QCS will need to be considered when integrating quantum noise affects into more finely tuned simulations.

\section{Demonstration}

The current era of quantum computing is commonly referred to as the Noisy Intermediate-Scale Quantum (NISQ) era. This is a result of quantum computers having a limited number of qubits and necessitating intricate error mitigation techniques that are beyond the scope of this paper. As we previously mentioned, we can take utilize the existing NISQ resources for testing the QCS framework in the presence of quantum noise. As a result, we choose to demonstrate the QCS framework on Amazon Braket, Amazon's cloud quantum computing platform \cite{braket}.

\subsection{Description of Model} \label{modeldesc}

LIDAR geoimaging represents a powerful technique for data analysis and data reconstruction, leveraging light detection and ranging technology to capture detailed spatial information. By emitting laser pulses and measuring the time it takes for the reflected light to return, LIDAR systems can create high-resolution, three-dimensional maps of landscapes and structures. This method has proven instrumental in various fields, such as topographic mapping, vegetation analysis, and urban planning \cite{microwave_image, WSN1}. The precise data obtained through LIDAR allows for accurate reconstruction of geographic features, enabling applications from environmental monitoring to infrastructure development. This makes it an ideal candidate for practical tests and allows for exploration of how LIDAR geoimaging facilitates data analysis and data reconstruction via compressive sensing, demonstrating its efficacy in producing detailed and reliable representations of complex environments. For the purposes of this work, the Concurrent Artificially-intelligent Spectrometry and Adaptive Lidar System (CASALS) satellite currently under development at NASA provides a practical LIDAR dataset that would be ideal for testing.

\begin{figure}[h]
    \centering
    \includegraphics[width=0.9\columnwidth]{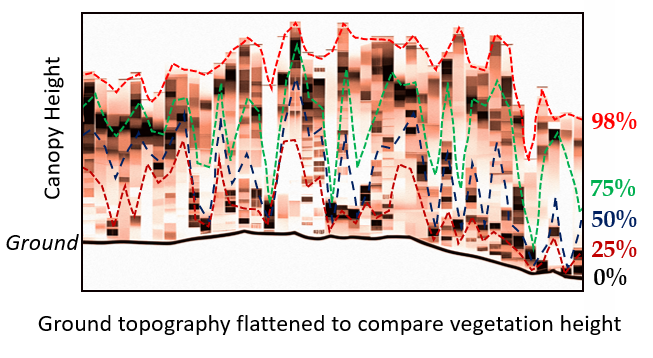}
    \caption{Visualization of how the LIDAR waveforms are split into five energy quartiles for a random segment of LIDAR data.}
    \label{fig:LIDAR}
\end{figure}

\begin{figure}[h]
    \centering
    \includegraphics[width=0.9\columnwidth]{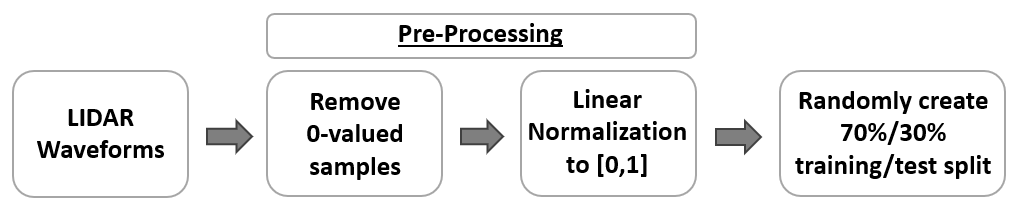}
    \caption{Steps for Data Processing}
    \label{fig:dataproc}
\end{figure}

The LIDAR waveforms for experiments in this work are represented via 5 pixels, each representing the energy quartile heights of the trees relative to the ground shown in the example of \figurename~\ref{fig:LIDAR}. This greatly reduces the dimensionality of the data while preserving key forest structure attributes. In order to pre-process the data, first all $0$-valued samples are removed and the remaining samples are linearly-normalized to a range of $[0, 1]$. We then randomly create a $70\% / 30\%$ training/test split from the $89,650$ remaining samples. \figurename~\ref{fig:dataproc} illustrates the steps for data preparation. All Born machines discussed in section \ref{modeldesc} are generated using samples randomly selected from the training split, whereas all projection trials in Section \ref{taining} use data selected from a small subset of the testing split.

\subsection{Training} \label{taining}

All Born machine circuits are generated using local, statevector simulations via Amazon Braket's python SDK. Due to Braket's inability to perform partial measurements, the resulting statevector is converted into a Qiskit object. The control qubits are then measured and upon successfully measuring all $0$s, the reduced density matrix of the superposition of samples $\psi_{ave}$ (``quantum average") is saved. The largest random subset that can be drawn from the training split features for our Born machine circuit is $2^{15}$ samples, hence the quantum average generated using this subset is termed $\psi_{global}$. In order to determine the smallest random subset that sufficiently represents $\psi_{global}$, we randomly create $5$ random independent training subsets of sizes $2^3$, $2^4$, up to $2^{13}$. Only three independent subsets can be generated of size $2^{14}$.

\begin{figure}[t]
    \centering
    \includegraphics[width=\linewidth]{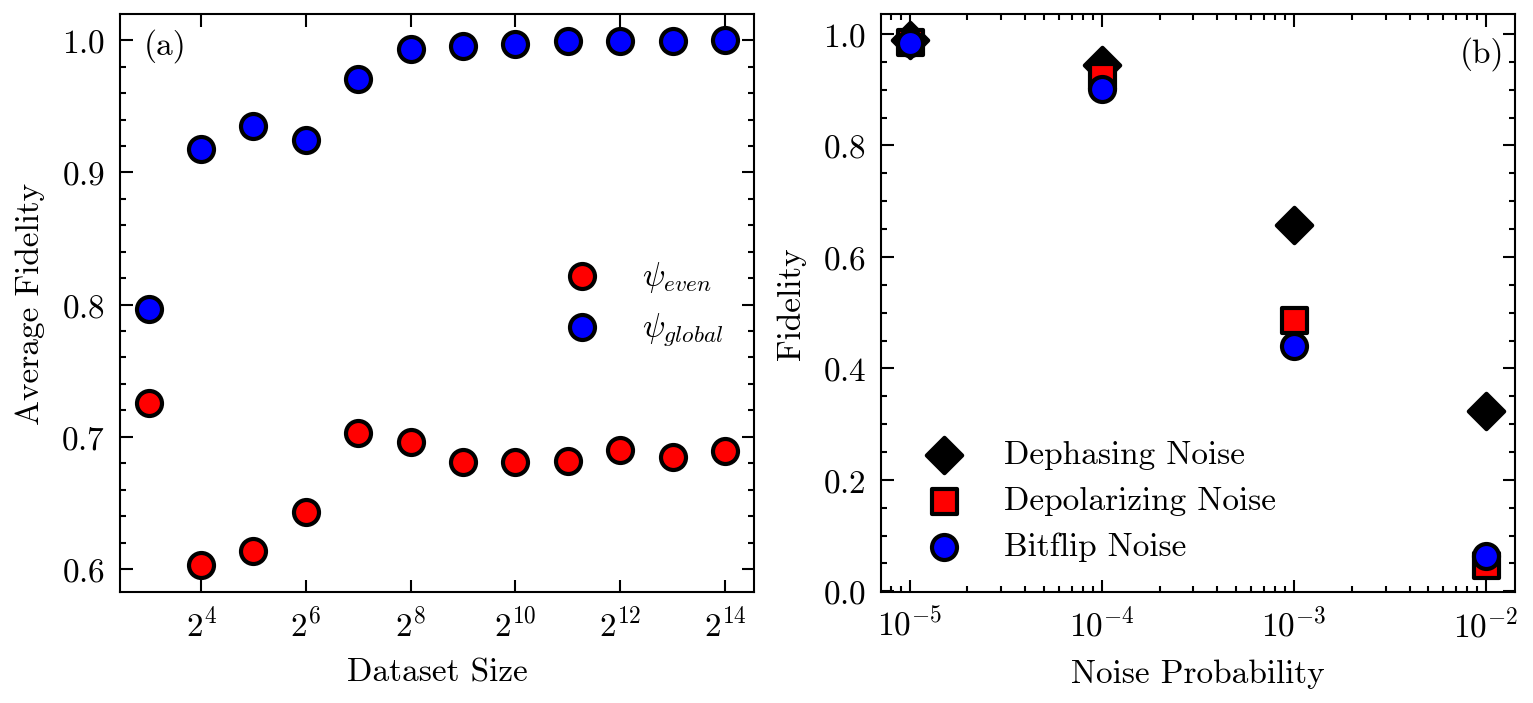}
    \caption{(a) Average fidelity of 5 datasets randomly selected from the training set with $\psi_{global}$ (blue) and $\psi_{even}$ (red) as a function of the dataset size. The average fidelity data with $\psi_{global}$ indicates that there is diminishing returns after randomly selecting $2^8$ (256) training samples. (b) Using a random dataset of 256 samples, plot the fidelity of the resulting quantum average with $\psi_{global}$ for dephasing noise (black diamond), depolarizing noise (red square), and bitflip noise (blue circle) as a function of noise probability. Here, the fidelity data with $\psi_{global}$ indicates that for a 5-pixel LIDAR dataset size of size 256, a noise probability on the order of $10^{-4}$ or smaller to reasonably preserve $\psi_{global}$.}
    \label{fig:training_results}
\end{figure}
 
In \figurename~\ref{fig:training_results}(a), we plot the average fidelity of these smallest training sets with $\psi_{global}$, which shows only $256$ randomly selected samples are necessary to create a quantum average with an average fidelity of at least $0.98$ with $\psi_{global}$. We also plot the average fidelity of these datasets with an equally-weighted quantum average of all possible basis states $\psi_{even}$ (or samples). This serves as a check to ensure that there is a meaningful uneven distribution of possible samples encoded into the quantum average.

Using one of the training subsets with $256$ samples, in \figurename~\ref{fig:training_results}(b) we examine the impact of aforementioned three models of quantum noise on our Born machines. Given that it is feasible we could control the unitary encodings for each sample on separate sets of control qubits or re-calibrate the qubits in-between samples, we only apply noise to the qubits that feed into each $\psi_{ave}$. For each type of noise described in sections Subsection \ref{flipnoise} through Subsection \ref{shotnoise}, we average $5000$ reduced density matrices generated via Braket statevector simulations. These matrices are then averaged to create a classical mixture that represents a noisy $\psi_{ave}$. Taking the fidelity of this with $\psi_{global}$ gives a measurement of information loss as a result of noise. Although dephasing noise seems to perturb the resulting quantum state less than bitflip and depolarizing noise, for all noise types the probability must be on the order of $10^{-4}$ or smaller to result in a noisy $\psi_{ave}$ that approximates $\psi_{global}$.

\subsection{Quantum Compressive Sensing with Quantum Noise}

Assuming we can generate a quantum average that approximates $\psi_{global}$ en masse, the next step is to build a circuit that decays away the samples that don't match with our classically measured signal. As discussed in Subsection~\ref{sec:projectionQITE}, QITE is the most flexible projection algorithm of the three proposed by QCS with respect to possible data-encoding schemes and training processes. We now present the QCS framework with QITE described in Subsection~\ref{sec:projectionQITE} featuring the noise processes outlined in Section~\ref{qnoise}. 

\begin{figure}
    \centering
    \includegraphics[width=\linewidth]{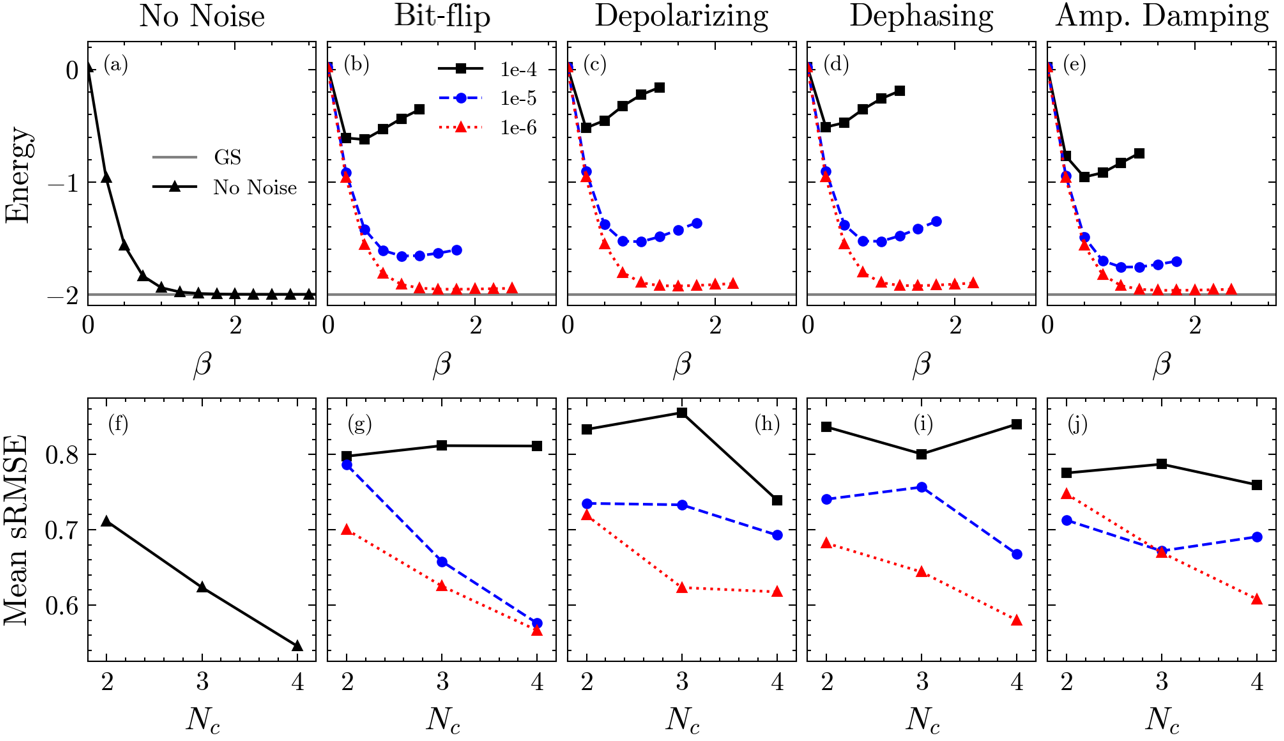}
    \caption{Demonstration of the QITE projection algorithm on a LIDAR dataset. (a) Convergence of noiseless QITE projection algorithm (black diamond) for the Hamiltonian $H=-(ZIIII + IIZII)$ to its ground state (GS). (b) Same as (a) except including single-gate bit-flip noise at probabilities of $1\mathrm{e}{-4}$, $1\mathrm{e}{-5}$, $1\mathrm{e}{-6}$ (black square, blue diamond, red triangle). (c)-(e) Same as (b) except for depolarizing, dephasing, and amplitude damping noise, respectively. (f) Mean per-pixel sRMSE, as defined in Equation~\ref{sRMSE}, for noiseless QCS algorithm using training set of $256$ samples and independent testing set of $64$ samples. (g)-(j) Same as (f) except for single-gate bit-flip noise at probabilities of $1\mathrm{e}{-4}$, $1\mathrm{e}{-5}$, $1\mathrm{e}{-6}$ (black square, blue diamond, red triangle). (h)-(j) being identical demonstrations for depolarizing, dephasing, and amplitude damping noise, respectively.}
    \label{fig:rmseplot}
\end{figure}

\begin{figure}
    \centering
    \includegraphics[width=\linewidth]{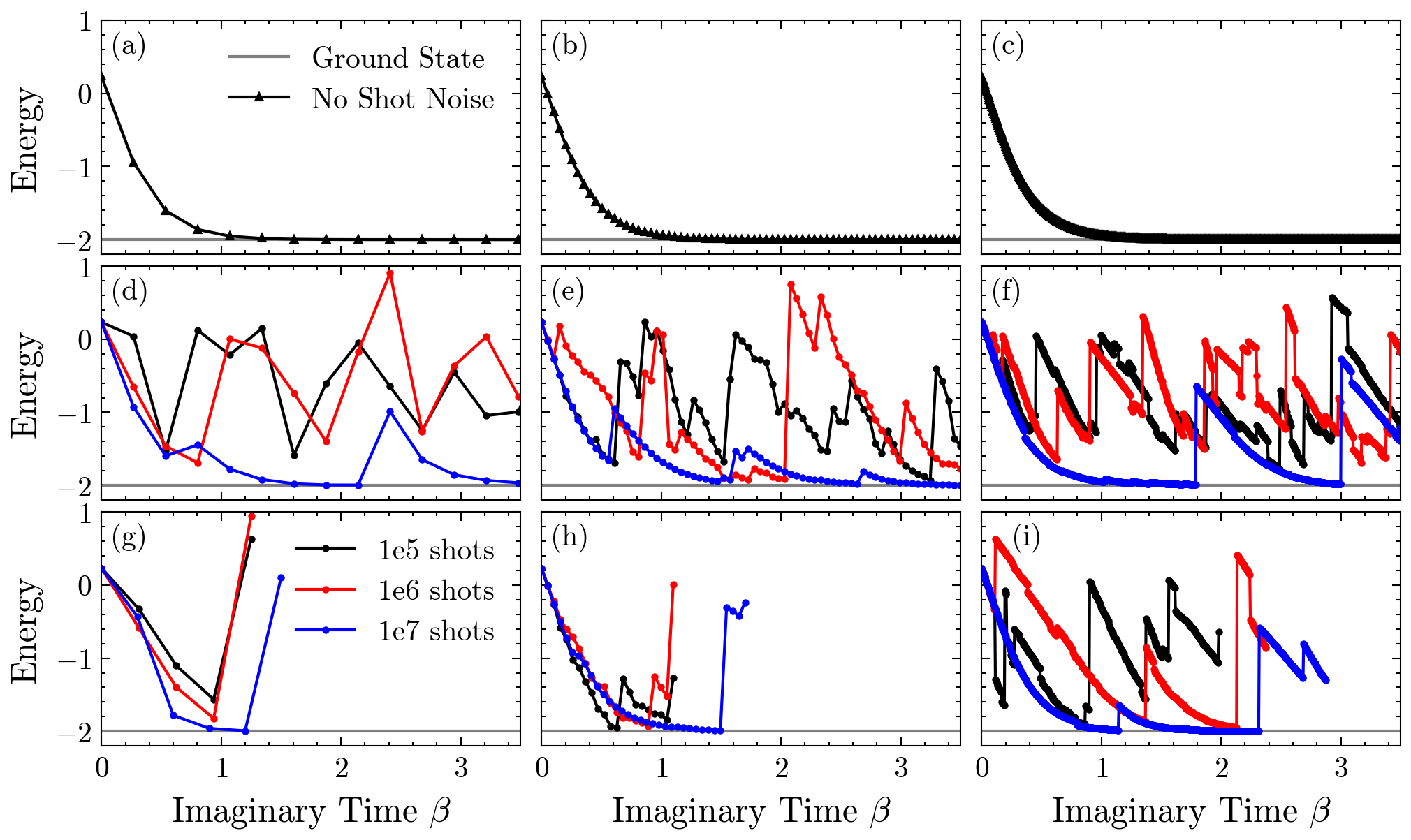}
    \caption{QITE runs of three different tomography approaches while varying the number of shots per observable and $d\beta$ for the 3-qubit Hamiltonian $H=-ZII - IIZ$. (a) QITE of $H$ using idealized tomography for $d\beta=0.3$ (black triangle). Ground state (grey) plotted at $-2$. (b) QITE of $H$ using idealized tomography for $d\beta=0.05$ (black triangle). (c) QITE of $H$ using idealized tomography for $d\beta=0.005$ (black triangle). (d) QITE for $H$ using 1e5 (black circle), 1e6 (red circle), and 1e7 (blue circle) shots per observable. (e) Same as (d) but for $d\beta=0.05$. (f) Same as (d) but for $d\beta=0.005$. (g) Same as (d) but discarding up to 30 failed QITE iterations before accepting an increase in Energy. (h) Same as (g) but for $d\beta=0.05$. (i) Same as (g) but for $d\beta=0.005$.} 
    \label{fig:shotnoise}
\end{figure}

Figure~\ref{fig:rmseplot} demonstrates the QITE projection algorithm on an exemplary LIDAR dataset in the presence of no noise, bit-flip noise, depolarizing noise, dephasing noise, and amplitude damping noise, where $N_c$ is the number of qubits measured classically. As a figure of merit for evaluating the performance in each noisy environment, we utilize the mean scaled root mean square error (sRMSE) metric defined as follows: 
\begin{equation} \label{sRMSE}
\text{Mean sRMSE} = \sqrt{\frac{\sum_{i} \left( \frac{P_i - R_i}{\sigma_{i}} \right)^2}{\upsilon}},
\end{equation}
where $i$ represents the indices being guessed, $P_i$ is the prediction, $R_i$ is the real value, $\upsilon$ is the number of guesses, and $\sigma_{i}$ is the standard deviation of the full training data of pixel $i$. It should be noted that the incorporation of $\sigma_i$ helps prevent any single pixel from dominating the error sum. Figure~\ref{fig:rmseplot}(a) illustrates a convergence of the noiseless QITE projection algorithm (i.e. black diamond) for the Hamiltonian $H=-(ZIIII + IIZII)$ to its ground state. Figure~\ref{fig:rmseplot}(b) is the same as Figure~\ref{fig:rmseplot}(a) except that it includes single-gate bit-flip noise at probabilities of $1\mathrm{e}{-4}$ (i.e. black square), $1\mathrm{e}{-5}$ (i.e. blue diamond), $1\mathrm{e}{-6}$ (i.e. red triangle), with Figures~\ref{fig:rmseplot}(c)-(e) being identical demonstrations for depolarizing, dephasing, and amplitude damping noise, respectively. Figure~\ref{fig:rmseplot}(f) plots the Mean sRMSE vs. $N_c$ for the noiseless QCS framework utilizing a training set size of $256$ samples and independent testing set of $64$ samples. Similarly, Figure~\ref{fig:rmseplot}(g) is Mean sRMSE vs. $N_c$ for a single-gate bit-flip noise at probabilities of $1\mathrm{e}{-4}$ (i.e. black square), $1\mathrm{e}{-5}$ (i.e. blue diamond), $1\mathrm{e}{-6}$ (i.e. red triangle), with Figures~\ref{fig:rmseplot}(h)-(j) being identical demonstrations for depolarizing, dephasing, and amplitude damping noise, respectively. It should be noted that pixel indices are randomly selected for $2$, $3$, $4$ classically measured pixels. Furthermore, the small $64$ testing samples introduces enough discrepancy between pixels to cause errors in some of the results. Figure~\ref{fig:rmseplot} also shows the scenario where the low sparsity and small size of the dataset can hinder many of QCS's strengths. Despite the errors caused by the limited number of testing samples and the dataset's disadvantages for QCS, Figures~\ref{fig:rmseplot}(g)-(j) still show promising trends in the Mean sRMSE, indicating that as more classical pixels are measured, the error per pixel decreases. Therefore, a future iteration of this demonstration with a larger and more sparse dataset should yield even more promising results. Of course, we acknowledge that noise, both in terms of strength and logical impact, will continue to play a key role in the performance of projection techniques like QITE, and by extension, the QCS framework as a whole.

Another type of noise, shot noise, is inherent source in quantum algorithms and is especially prominent in QITE as each iteration features quantum state tomography. Figure~\ref{fig:shotnoise} displays the results for QITE runs of three different tomography approaches, adjusting both the number of shots per observable and $d\beta$ for the 3-qubit Hamiltonian $H=-ZII - IIZ$. Figure~\ref{fig:shotnoise}(a) is QITE of $H$ using idealized tomography for $d\beta=0.3$, while Figure~\ref{fig:shotnoise}(b) is QITE of $H$ using idealized tomography for $d\beta=0.05$, and Figure~\ref{fig:shotnoise}(c) is QITE of $H$ using idealized tomography for $d\beta=0.005$. Subfigures~\ref{fig:shotnoise} (d)-(f) illustrate QITE for $H$ using different shots per observable for varying $d\beta$. Finally, Subfigures~\ref{fig:shotnoise} (g)-(i) also illustrate QITE for $H$ using different shots per observable for varying $d\beta$, but this time discard up to 30 failed QITE iterations before accepting an increase in Energy. Even for a low-qubit $H$ with real coefficients and fully-commuting terms, shot noise is very prominent in the framework. Furthermore, implementing a check for highly-perturbative instances of shot-noise and choosing a medium-sized $d\beta$ as in Figure~\ref{fig:shotnoise}(i) significantly mitigates the impact of shot noise of $H$ on even the lowest number of shots per observable. 

 \vspace{-0.22in}

\section{Long Term Possible Directions and Applications} \vspace{-0.18in}

Given the results of the current QCS pipeline, we will now outline potential next phases of its development. This section explores two potential research directions aimed at evolving QCS into a practical hybrid quantum-classical pipeline for signal processing.

\subsection{Compressive Sensing}

As previously discussed, compressive sensing is a classical sensing protocol that leverages known sparsity in signal sources to achieve sampling rates below the Nyquist rate \cite{CS1, CS2, CS3, CS4}. The central idea of compressive sensing is that it allows for the accurate recovery of sparse signals from far fewer measurements than traditional methods by exploiting their inherent simplicity \cite{CS2, CS3}. There are typically two types of approaches to achieve this: convex $l_1$ minimization as an approximation of the $l_0$ problem or a greedy algorithm that iterates over the reduced measurements and finds the k-largest coefficients in a known sparse basis with some probability \cite{sfft}. Compressive sensing has practical applications in various fields, significantly impacting data acquisition and reconstruction processes. In medical imaging, it accelerates MRI scans by reducing the number of required measurements while maintaining high image quality \cite{MRI}. In wireless communications, it enhances signal recovery and reduces bandwidth usage, which is crucial for efficient data transmission \cite{CS4}. Additionally, compressive sensing is used in astronomical imaging to improve the resolution of telescopic observations while minimizing data collection time \cite{candes_sparsity}. 

Although the QCS framework was partly inspired by compressive sensing, there is still room for further integration of compressive sensing techniques. A natural direction for further research is to explore large-data applications with suitable sparsity structures. There are two potential approaches to achieve this with the current algorithms. The first approach involves classically sampling from the training set at a reduced rate and using classical compressive sensing to reconstruct the full training set in a sparse basis. The reconstructed sparse training set can then be loaded onto quantum hardware for projection. Additionally, since the quantum average will be a superposition of sparse signals, integrating compressive sensing into the QCS QITE implementation has the potential to accelerate the tomography steps \cite{Gross2010}. Alternatively, depending on the computational cost of classical compressive sensing for a given dataset, another approach is to hybridize the signal reconstruction process to reduce data preprocessing. Instead of loading the reconstructed training samples in superposition, this method involves loading the reduced measurements of the training samples in superposition and performing $l_0$ minimization on quantum hardware. A promising note is that a similar algorithm, known as the quantum fast Fourier transform (qFFT), has been successfully developed to perform the fast Fourier transform on a superposition of signals in basis encoding \cite{qfft}. Given the structure of both the qFFT and our desired quantum average, probabilistic greedy compressive sensing approaches are likely the most suitable for this scheme.

\subsection{Data Encoding}

Moving forward, it is likely that the current encoding will need to be specifically adapted for both signal reconstruction and generation applications. Even with QCS's relatively small and idealized algorithms, optimal performance within the framework can only be achieved by customizing the entire mapping to the LIDAR dataset. Consider the following simplistic example in which the training set features two distinct signals at equal probability $w_1 = [0, 1, 0]$ and $w_2 = [0, 0.5, 1]$ and we let the midpoint p of each pixel be $0.5$. Assuming we classically measure pixels $1$ and $2$ as $[0, 1]$ looking to reconstruct $3$, even our best projection algorithm, QITE, will only decay off the portion of $w_2$ that overlaps with $w_1$. Approximately $33 \%$ of the time after we have run QITE, then, we will incorrectly reconstruct a $1$ despite a clear dividing line on pixel $2$. However, in applying a re-scaling mapping by setting $v=[0, 0.75]$, this error rate is decreased to below $10 \%$. Furthermore, the error rate can be eliminated completely by either adding an additional qubit for pixel $2$ to set $0.5$ as an orthogonal state to $0$ and $1$ or in pre-processing the dataset such that $0.5$ maps to $\ket{0}$ for pixel two.

 With that in mind, it will be necessary to define the specifics of problem and dataset structures for further pursuit of this research direction. From there, classical pre-processing techniques based on training-set statistics can be merged with encoding techniques to tune optimal mapping of data throughout the pipeline. For a training procedure similar to our current circuit and encoding scheme, four fundamental methods of data encoding are available: basis encoding, amplitude encoding, angle encoding (QCS's current method), and Hamiltonian encodings \cite{Schuld2021}. In addition, tunable parameters can be introduced to QCS's framework similar to the proposed methods for optimizing image encodings for feature extraction \cite{Wang_2022}. Specifically, a framework that adjusts encoding schemes to better capture relevant visual information and improve downstream image analysis tasks is utilized with fine-tuning encoding strategies to address specific image characteristics and noise levels. This leads to more accurate and robust image processing results and demonstrates significant improvements in performance for various image processing applications compared to traditional encoding techniques. 
 
\vspace{-0.22in}

\section{Conclusion}

 This work presents a practical implementation of QCS on Amazon Braket, demonstrating its potential for advancing quantum data-driven approaches to compressive sensing. By leveraging the QITE technique, we explore the framework's capabilities under quantum noise and discuss possible future directions. The current QCS framework provides a strong foundation for various long-term applications and has room for growth as quantum computing challenges are addressed theoretically and practically.

 \section{Acknowledgments}
 The authors would like to thank Dr. Erika Jones for useful discussions, which helped us significantly in clarifying the ideas of this work

\vspace{-0.22in}


\newpage
\bibliographystyle{IEEEtran}
\bibliography{QCS_Paper.bib}


\end{document}